\documentclass[fleqn,10pt]{wlscirep}
\usepackage[utf8]{inputenc}
\usepackage[T1]{fontenc}
\usepackage{lmodern}  
\usepackage{graphicx}
\usepackage{makecell}
\usepackage{xcolor}
\usepackage{subcaption}
\usepackage{hyperref}
\usepackage[table]{xcolor} 

\newcommand{\datasetname}{World-POI} 
\newcommand{\red}[1]{\color{red}{ERROR HERE}}
\newcommand{\reffig}[1]{Figure~\ref{#1}}
\newcommand{\reftab}[1]{Table~\ref{#1}}

\newcommand{\qgismapcopyright}{The map was generated in QGIS using the ESRI Physical Map basemap layer. }
\newcommand{\pymapcopyright}{The map was generated in Python and the source code is on GitHub}
\newcommand{\githuburl}{\url{https://github.com/onspatial/world-poi}}
\newcommand{\osfurl}{\url{https://osf.io/p96uf}}

\newcolumntype{C}[1]{>{\centering\arraybackslash}m{#1}}

\title{\datasetname{}: Global Point-of-Interest Data Enriched from Foursquare and OpenStreetMap as Tabular and Graph Data}

\author[1,*]{Hossein Amiri}
\author[1]{Mohammad Hashemi}
\author[1,*]{Andreas Z{\"u}fle}
\affil[1]{Department of Computer Science and Informatics, Emory University Atlanta, USA}
\affil[*]{Corresponding authors: Hossein Amiri (hossein.amiri@emory.edu) and Andreas Z{\"u}fle (azufle@emory.edu) }

\begin{abstract}
Recently, Foursquare released a global dataset with more than 100 million points of interest (POIs), each representing a real-world business on its platform. However, many entries lack complete metadata such as addresses or categories, and some correspond to non-existent or fictional locations. In contrast, OpenStreetMap (OSM) offers a rich, user-contributed POI dataset with detailed and frequently updated metadata, though it does not formally verify whether a POI represents an actual business.
In this data paper, we present a methodology that integrates the strengths of both datasets: Foursquare as a comprehensive baseline of commercial POIs and OSM as a source of enriched metadata. The combined dataset totals approximately 1 TB. While this full version is not publicly released, we provide filtered releases with adjustable thresholds that reduce storage needs and make the data practical to download and use across domains. We also provide step-by-step instructions to reproduce the full 631 GB build.
Record linkage is achieved by computing name similarity scores and spatial distances between Foursquare and OSM POIs. These measures identify and retain high-confidence matches that correspond to real businesses in Foursquare, have representations in OSM, and show strong name similarity.
Finally, we use this filtered dataset to construct a graph-based representation of POIs enriched with attributes from both sources, enabling advanced spatial analyses and a range of downstream applications.

\end{abstract}
\begin{document}

\flushbottom
\maketitle

\thispagestyle{empty}

\begin{figure}
	\centering
	\includegraphics[width=\linewidth]{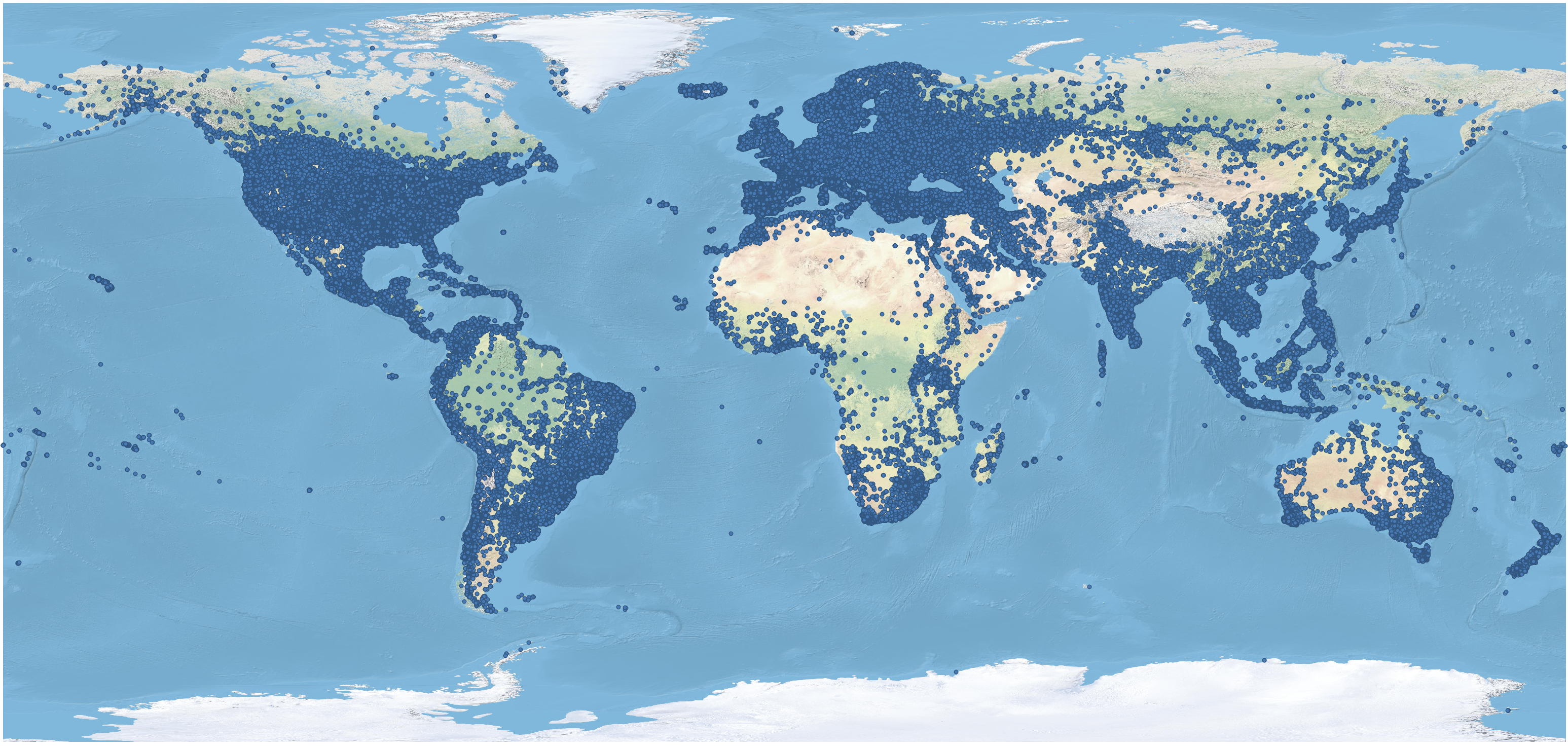}
	\caption{Visualization of a representative sample output from the \datasetname{} database, filtered to include entries with Levenshtein name similarity scores greater than 0.5. \qgismapcopyright}
	\label{fig:fosm_overview_map}
\end{figure}

\section*{Background \& Summary}
Points of Interest (POI) datasets are foundational for a wide range of spatial computing applications, including urban analytics~\cite{kandt2021smart,ommi2024machine}, mobility modeling and simulation~\cite{amiri2023massive,amiri2024geolife+}, place recommendation~\cite{saiph2012m,takeuchi2006cityvoyager,hashemi2025placefm,hashemi2025points}, anomaly detection~\cite{zhang2024large,zhang2024transferable,amiri2024urban}, modeling infectious disease spread~\cite{kohn2023epipol}, and enabling location-based services~\cite{amiri2024patterns}. As these applications continue to evolve, the demand for comprehensive high-quality POI datasets has intensified. However, existing POI datasets often exhibit trade-offs between spatial coverage, semantic richness, and cost, limiting their utility for certain tasks.

There are several publicly available POI datasets, ranging from commercial to open-source options. Prominent paid sources include the Google Places API~\cite{googleplace}, ADVAN~\cite{advanresearch}, Bing Maps~\cite{bingmaps}, HERE Maps~\cite{heremaps}, and Precisely~\cite{precisely}. These services typically offer extensive global coverage, rich metadata, and verified business listings, but come with licensing fees or usage-based pricing models. For example, the Google Places API provides access to over 200 million POIs but incurs substantial costs at scale. In contrast, open-source and freely available datasets such as Foursquare~\cite{foursquare}, OpenStreetMap (OSM)~\cite{osm}, the USGS Geographic Names Information System (GNIS)~\cite{gnis}, and Overture Maps~\cite{overture} offer cost-free access, though with varying levels of completeness, consistency, and quality. Among these, Foursquare and OSM are the focus of this work, and we validate our integrated dataset using Google Maps as an external reference.

Commercial datasets such as Foursquare generally provide high-precision, verified business entries and up-to-date information. However, they often lack consistent or complete metadata, including standardized category taxonomies, administrative boundaries, or contextual geographic attributes~\cite{foursquare}. Furthermore, user-generated contributions may introduce inaccuracies or erroneous points of interest (POIs) into the database. In contrast, OSM offers a comprehensive, community-driven platform with rich semantic annotations, flexible tagging schemes, and broad spatial coverage. However, its crowd-sourced nature means that business verification is informal, and metadata can sometimes be inconsistent, incomplete, or outdated~\cite{osm}.

In this paper, we address the complementary limitations of these sources by introducing \textbf{\datasetname{}}, an enriched and integrated POI dataset that aligns and merges records from Foursquare and OSM. Using a hybrid matching approach based on spatial proximity and name similarity, we construct a high-confidence mapping of POI pairs that combines the precision and reliability of Foursquare with the semantic depth and geographic richness of OSM. The resulting dataset provides detailed metadata, standardized category labels, and accurate spatial annotations. An example of a representative sample from the \datasetname{} database is shown in \reffig{fig:fosm_overview_map}, illustrating entries filtered by Levenshtein name similarity scores greater than 0.5. In other words, if a POI appears in both Foursquare and OSM, shares similar spatial attributes, and exhibits high name similarity, it is more likely to represent the same real-world location. Moreover, POIs present in both datasets are more likely to correspond to actual businesses rather than user-generated or fictitious entries added to gain contribution points within the platform.

\textbf{\datasetname{}} is released in both tabular and graph-based formats to support diverse analytical workflows, such as geographic knowledge graph construction, spatial clustering, and simulation-based modeling. By bridging the strengths of two widely used POI sources, \datasetname{} enables more accurate modeling, analysis, and simulation of real-world spatial environments. This dataset serves as a high-utility resource for researchers and practitioners in urban computing, geoinformatics, and spatial data science.

\section*{Methods}
\begin{figure}
	\centering
	\includegraphics[width=1\linewidth]{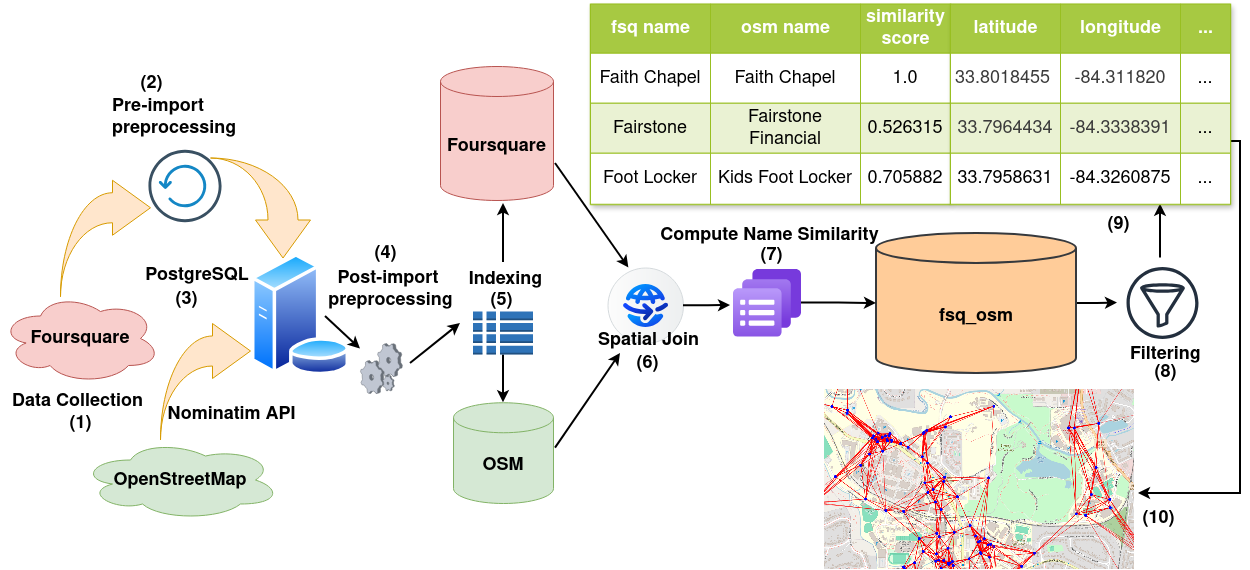}
	\caption{Overview of the pipeline for constructing the \datasetname{} dataset by merging Foursquare and OpenStreetMap POIs through preprocessing, spatial indexing, and similarity-based filtering. \pymapcopyright}
	\label{fig:pipeline}
\end{figure}

In this section, we outline the methodology used to construct the \datasetname{} dataset, ensuring both transparency and reproducibility. The complete workflow is illustrated in \reffig{fig:pipeline}. In summary: (1) collect POI data from Foursquare and OpenStreetMap (OSM); (2) perform data cleaning and initial preprocessing on the Foursquare dataset; (3) import the cleaned Foursquare data and downloaded OSM data into PostgreSQL/PostGIS; (4) conduct additional preprocessing to harmonize attributes and resolve inconsistencies; (5) create spatial indexes; (6) perform spatial joins to identify candidate POI pairs based on geographic proximity; (7) compute name-similarity scores between Foursquare and OSM entries; (8) apply similarity thresholds to retain high-confidence matches; (9) generate tabular outputs; and (10) construct corresponding graph-based representations.

\subsection*{Step (1) — Data Collection}

To obtain the Foursquare data, we accessed Foursquare’s publicly available cloud storage (hosted on Amazon S3) and downloaded a comprehensive collection of place-related datasets. These datasets were originally provided in split, compressed Parquet format and organized by category information and place data to facilitate access.

For OpenStreetMap (OSM) data, we downloaded the full dataset directly from the official OSM website. To make the data suitable for analysis and querying, we utilized Nominatim, a geocoding and data conversion tool, to import the raw OSM data into a PostgreSQL-compatible SQL format. This process generated multiple relational tables, among which the places table—containing detailed information about points of interest (POIs)—served as the primary input for integration.

\subsection*{Step (2) — Pre-import Preprocessing}
After downloading the Foursquare data, the Parquet files were converted into a more accessible and widely supported format (CSV). All converted files were then aggregated into a single unified dataset. During this process, geometry values were removed, numerical fields were converted to numeric types, and string attributes were enclosed in quotation marks (“”) to ensure proper parsing. This cleaning step ensured consistent formatting and completeness prior to database ingestion. In addition, columns were renamed to maintain a consistent naming convention by prefixing each with “fsq\_” if it did not already begin with that prefix. These steps ensured that the dataset was fully standardized and ready for structured ingestion into PostgreSQL/PostGIS.

\subsection*{Step (3) — Data Import into PostgreSQL/PostGIS}

The cleaned Foursquare dataset was imported into a PostgreSQL database with the PostGIS extension enabled to support spatial operations. A new database, \texttt{fsq-osm}, was created to store both the Foursquare and OpenStreetMap (OSM) datasets in structured form.
For the Foursquare data, a dedicated table (\texttt{foursquare}) was created following the schema defined in our GitHub repository~\cite{github}. The data were loaded using the \texttt{\textbackslash copy} command, which efficiently imports large CSV files while preserving headers, delimiters, and quoted strings.
The OSM data, downloaded in PBF format, were first processed using the Nominatim import tool to convert the raw OSM file into a PostgreSQL-compatible database. 

\subsection*{Step (4) — Post-import Preprocessing}
Following the data import, additional preprocessing was conducted to harmonize attributes and resolve inconsistencies between the Foursquare and OSM schemas. This included aligning coordinate systems, validating geometries, and removing duplicate or corrupted records.
From the imported OSM database, we used the place table and only features containing valid \textit{name} attributes and not classified as \textit{highways} were extracted and exported as a filtered CSV file. During this step, all column names were prefixed with \texttt{osm\_} to maintain a consistent naming convention. In addition, the centroid of each original geometry was calculated to obtain representative latitude and longitude coordinates. The filtered data were subsequently exported to a CSV file and imported into the \texttt{osm} table within the \texttt{fsq-osm} database using the \texttt{\textbackslash copy} command.
Then, Geometry columns (\texttt{fsq\_geom} and \texttt{osm\_geom}) were generated for both \texttt{foursquare} and \texttt{osm} tables  using PostGIS functions that convert longitude and latitude values into point geometries, standardized under the WGS 84 coordinate reference system (EPSG:4326) to ensure consistency.

\subsection*{Step (5) — Spatial Indexing}

To optimize performance for spatial queries and joins, GiST (Generalized Search Tree) indexes were created on the geometry columns of both datasets (\texttt{fsq\_geom} and \texttt{osm\_geom}). These spatial indexes substantially improved query efficiency, reducing computation time during spatial matching and distance calculations.

\subsection*{Step (6) — Spatial Join Based on Geographic Proximity}

To align and integrate the two data sources, a spatial join was performed between Foursquare POIs (\texttt{foursquare} table) and OSM POIs (\texttt{osm} table) based on geographic proximity. For each Foursquare record, the nearest OSM feature within a 50-meter radius was identified. This threshold was chosen to balance precision and recall—accurately linking spatially corresponding entries while minimizing false matches and to limit the output size to a manageable volume (approximately 1 TB) for storage and further processing. In this step, the geographic distance between each pair was calculated and stored as a column in degrees.
The resulting integrated dataset combined the commercial metadata of Foursquare with the semantic richness of OSM, enabling comprehensive spatial and contextual analysis. 

\subsection*{Step (7) — Name Similarity Computation}

For each spatially matched POI pair, name similarity between Foursquare and OSM entries was computed using two complementary text-matching methods: trigram-based similarity and the Levenshtein~\cite{yujian2007normalized} distance metric. These measures quantified the textual closeness of place names, complementing spatial proximity as an additional criterion for validating matches. The resulting similarity values were stored in two separate columns—one for trigram-based scores and one for Levenshtein-based scores—within the integrated dataset to facilitate downstream filtering and confidence assessment.

\subsection*{Step (8) — High-Confidence Match Selection}

To enhance dataset quality, only high-confidence matches were retained based on a Levenshtein name-similarity threshold of 0.5. Pairs meeting or exceeding this threshold were considered reliable alignments between Foursquare and OSM entries. The resulting dataset, stored as \texttt{fsq\_osm\_filtered\_5\_lev}, contains harmonized POI records with unique \texttt{place\_id} identifiers from Foursquare and corresponding \texttt{osm\_id} values from OSM, forming a consistent and high-quality integrated dataset.

\subsection*{Step (9) — Tabular Dataset Generation}

After filtering, the integrated dataset was further refined for tabular export. Each record includes validated spatial coordinates, standardized metadata, name-similarity scores, and computed geographic distances. 
In addition, we assign a unique \texttt{poi\_id} as the primary key to the filtered dataset to handle potential duplicates of\texttt{ fsq\_place\_id}. This situation can occur when a single Foursquare place ID has multiple high-similarity matches in nearby areas—for example, Emory University, Emory University Hospital, and Emory University Library.
The finalized dataset was exported in CSV format to ensure broad compatibility with analytical and visualization tools. This structured output serves as the foundation for subsequent geospatial and semantic analyses. The resulting data are illustrated in \reffig{fig:fosm_overview_map}.

\subsection*{Step (10) — Graph-Based Representation}

In addition to the tabular dataset, a graph-based representation of \datasetname{} was generated. Each POI was modeled as a node connected to its $N$ nearest neighbors based on geographic proximity. Edge weights correspond to the spatial distances between nodes, capturing local spatial relationships and neighborhood structure. This graph representation facilitates network-based analyses of urban topology, connectivity, and spatial clustering. An example of the resulting graph for $N=10$ in the vicinity of Emory University is shown in \reffig{fig:graph_example_emory}.

\section*{Data Records}
In this section, we provide a detailed overview of the records contained in the dataset. The tabular format includes all available data fields, whereas the graph representation contains only the place identifiers and the distances between them. To obtain detailed information about a specific place identifier in the graph, the corresponding record from the tabular dataset should be referenced. In addition, we provide multiple versions of the dataset generated using different similarity thresholds, allowing users to select the one most suitable for their specific applications. However, throughout this paper, examples and analyses are based on the dataset filtered using a Levenshtein-based name-similarity threshold greater than 0.5. A complete list of all generated datasets, along with their corresponding thresholds and download links, is available in the project’s GitHub repository (\githuburl) and available for download at (\osfurl)

\begin{table}
	\centering
	\caption{Descriptions of columns in the \datasetname{} dataset.}
	\label{tab:data_record}
	\footnotesize
	\begin{tabular}{|m{0.15\linewidth}|m{0.40\linewidth}|m{0.25\linewidth}|}
		\hline
		\textbf{Column Name}   & \textbf{Description}                                         & \textbf{Example Value}                                                            \\ \hline
		fsq\_place\_id         & Unique identifier of the place (Foursquare)                  & 4a4a0fd6f964a52087ab1fe3                                                          \\ \hline
		fsq\_name              & Name of the place (Foursquare)                               & Starbucks                                                                         \\ \hline
		fsq\_latitude          & Latitude coordinate (Foursquare)                             & 33.768107                                                                         \\ \hline
		fsq\_longitude         & Longitude coordinate (Foursquare)                            & -84.34941113                                                                      \\ \hline
		fsq\_address           & Street address (Foursquare)                                  & 506 Moreland Ave NE                                                               \\ \hline
		fsq\_locality          & Neighborhood or locality (Foursquare)                        & Birmingham                                                                        \\ \hline
		fsq\_region            & Region, state, or province (Foursquare)                      & GA                                                                                \\ \hline
		fsq\_postcode          & Postal or ZIP code (Foursquare)                              & 30307                                                                             \\ \hline
		fsq\_admin\_region     & Higher-level administrative division (Foursquare)            & England                                                                           \\ \hline
		fsq\_post\_town        & Post town (Foursquare)                                       & Birmingham                                                                        \\ \hline
		fsq\_po\_box           & PO Box number (Foursquare)                                   & P.O. Box 41404                                                                    \\ \hline
		fsq\_country           & Country name (Foursquare)                                    & GB                                                                                \\ \hline
		fsq\_date\_created     & Date the record was created (Foursquare)                     & 2009-06-30                                                                        \\ \hline
		fsq\_date\_refreshed   & Date the record was last updated (Foursquare)                & 2025-06-22                                                                        \\ \hline
		fsq\_date\_closed      & Date the place was closed, if applicable (Foursquare)        & 2025-01-01                                                                        \\ \hline
		fsq\_tel               & Telephone number (Foursquare)                                & (404) 230-9085                                                                    \\ \hline
		fsq\_website           & Website URL (Foursquare)                                     & http://starbucks.com                                                              \\ \hline
		fsq\_email             & Email address (Foursquare)                                   & info@starbucks.com                                                                \\ \hline
		fsq\_facebook\_id      & Facebook page identifier or URL (Foursquare)                 & 22092443056.0                                                                     \\ \hline
		fsq\_instagram         & Instagram handle or URL (Foursquare)                         & starbucks                                                                         \\ \hline
		fsq\_twitter           & Twitter handle or URL (Foursquare)                           & starbucks                                                                         \\ \hline
		fsq\_category\_ids     & List of category IDs (Foursquare)                            & ['4bf58dd8d48988d1e0931735']                                                      \\ \hline
		fsq\_category\_labels  & Human-readable category names (Foursquare)                   & ['Dining and Drinking > Cafe, Coffee, and Tea House > Coffee Shop']               \\ \hline
		fsq\_placemaker\_url   & URL to Placemaker or API resource (Foursquare)               & \url{https://foursquare.com/placemakers/review-place/4a4a0fd6f964a52087ab1fe3   } \\ \hline
		fsq\_unresolved\_flags & Quality issues reported for a POI (Foursquare)               & ['duplicate']                                                                     \\ \hline
		fsq\_bbox              & Bounding box of the place (Foursquare)                       & \makecell{\{'xmin': -84.34941113,                                                 \\'ymin': 33.768107,\\'xmax': -84.34941113,\\'ymax': 33.768107\}}                                         \\ \hline
		fsq\_geom              & Point geometry (latitude/longitude) in WGS84 (Foursquare)    & 01010..1E24040                                                                    \\ \hline
		osm\_id                & Unique identifier of the OSM object                          & 1237615380                                                                        \\ \hline
		osm\_class             & Main category or class (OSM)                                 & amenity                                                                           \\ \hline
		osm\_type              & More specific type (OSM)                                     & cafe                                                                              \\ \hline
		osm\_name              & Name of the object (OSM)                                     & Starbucks                                                                         \\ \hline
		osm\_address           & Address or location description (OSM)                        & \makecell{"city"=>"Little Five Points                                             \\Village", "state"=>"GA",\\"street"=>"Moreland Avenue",\\"postcode"=>"30307",\\"housenumber"=>"506"} \\ \hline
		osm\_extratags         & Additional tags or metadata (OSM)                            & \makecell{"phone"=>"+1 404-230-9085",                                             \\"branch"=>"Little Five Points",\\"cuisine"=>"coffee\_shop",\\"takeaway"=>"yes",...}                 \\ \hline
		osm\_geometry          & Geometry of the object (OSM)                                 & 01010..1E24040                                                                    \\ \hline
		osm\_latitude          & Latitude in decimal degrees derived from osm\_geometry       & 33.7680684                                                                        \\ \hline
		osm\_longitude         & Longitude in decimal degrees derived from osm\_geometry      & -84.3494503                                                                       \\ \hline
		osm\_geom              & Point geometry derived from osm\_latitude and osm\_longitude & 01010..1E24040                                                                    \\ \hline
		\makecell{fsq\_osm\_name\_                                                                                                                                                \\similarity\_score\_trg} & Trigram-based name similarity                                & 0.99                                                                                                                           \\ \hline
		\makecell{fsq\_osm\_name\_                                                                                                                                                \\similarity\_score\_lev} & Levenshtein-based name similarity                            & 0.99                                                                                                                           \\ \hline
		fsq\_osm\_distance     & Distance between Foursquare and OSM locations in degrees                & 5.499317140236702e-05                                                             \\ \hline
	\end{tabular}
\end{table}

\subsection*{Tabular Data Structure}

The fsq\_osm table constitutes the final integrated dataset that merges detailed place information from Foursquare and OpenStreetMap (OSM). Each row in the table corresponds to a point-of-interest (POI), enriched with spatial, semantic, and contextual metadata from both sources.

\reftab{tab:data_record} lists the columns of the provided dataset along with their descriptions. Columns originating from Foursquare are prefixed with fsq\_, while those from OpenStreetMap are prefixed with osm\_.
Foursquare-derived fields include unique identifiers, place names, latitude and longitude coordinates, street address components, city, state, country, postal code, contact details (e.g., phone number, website), and associated social media handles (e.g., Twitter, Facebook). In addition, each record contains one or more categorical tags describing the place type or function. The spatial location is stored using a POINT geometry in WGS 84 (EPSG:4326) format, which enables efficient geospatial queries using PostGIS or other spatial databases.
OSM-derived fields enrich these records with community-curated metadata, including OSM identifiers, POI names, classification types (e.g., amenity, tourism, leisure), administrative levels, and freeform tag-value pairs contributed by mappers. These fields offer rich geographic and semantic context, such as whether the place is within a park, located on a university campus, or tagged with specific amenities. 
We provide two geometries for each OSM record: the original geometry from the OSM dataset and a point geometry calculated from its centroid. The centroid-based geometry ensures consistency with the Foursquare point geometries, aligned with the same spatial reference system, enabling uniform spatial operations between the two datasets.
To facilitate transparent integration and downstream filtering, we include three computed fields:

\begin{itemize}
	\item \textbf{fsq\_osm\_distance:} the great-circle distance (in degrees) between the Foursquare and OSM POI coordinates.
	
	\item \textbf{name\_similarity\_score\_trg:} a normalized trigram-based string similarity metric that quantifies the semantic resemblance between the names of the matched POIs.
	
	\item \textbf{name\_similarity\_score\_lev:} a normalized Levenshtein-based string similarity metric that quantifies the semantic resemblance between the names of the matched POIs.
\end{itemize}

We note that using these fields, filtering can be performed based on different thresholds and criteria. For instance, candidate pairs can be selected using a minimum name-similarity score (e.g., \texttt{name\_similarity\_score\_lev} $\geq$ 0.5) and a maximum spatial distance (e.g., \texttt{fsq\_osm\_distance} $\leq$ 0.001 degrees).  
Additionally, other similarity metrics or attributes can be computed and appended to the join table to enhance matching accuracy. However, the calculation and integration of these additional metrics are beyond the scope of this paper and are left for users or future work.

\subsection*{Graph Data Structure}
\begin{figure}[t]
	\centering
	\includegraphics[width=0.99\linewidth]{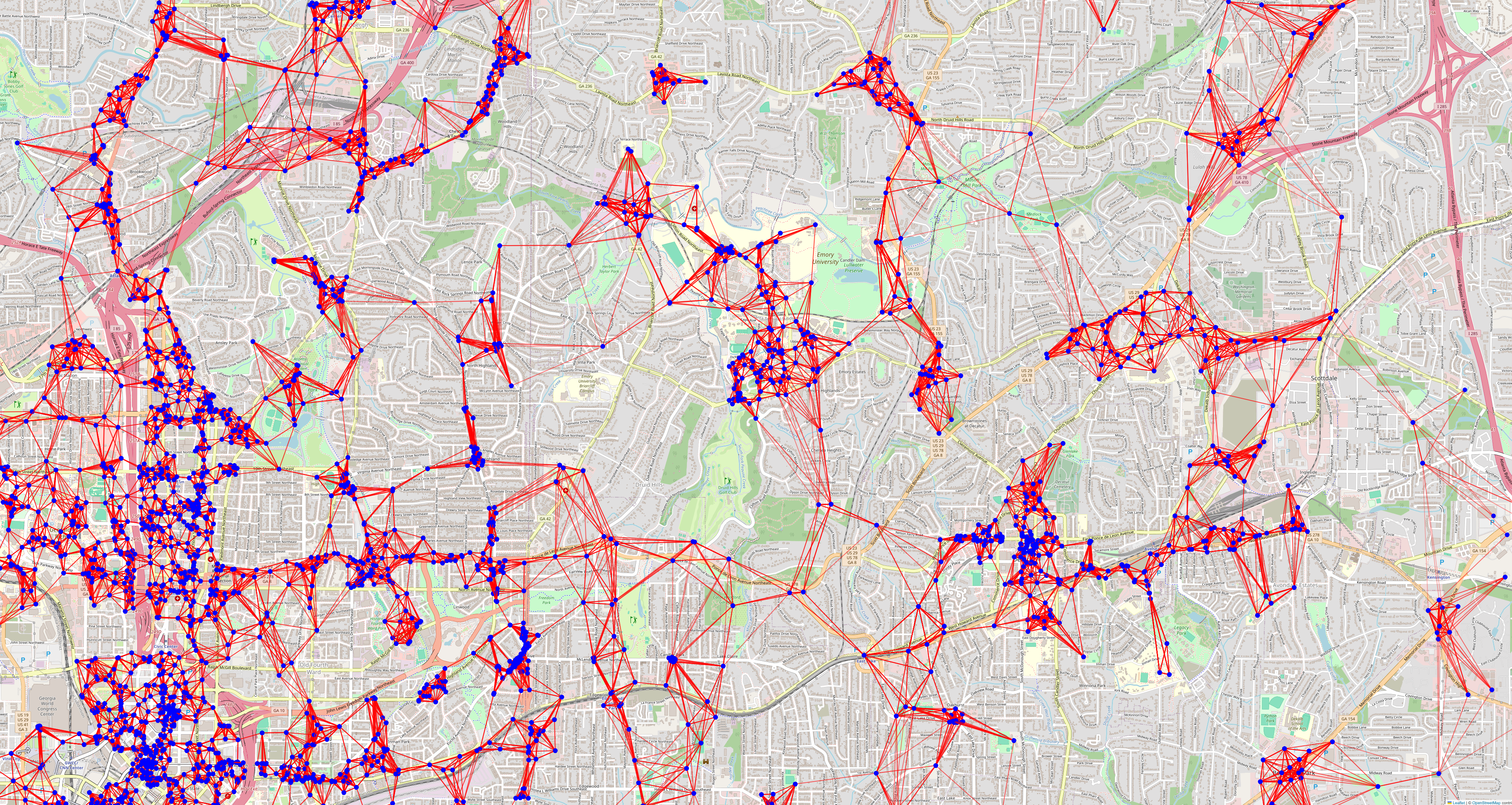}
	\caption{Visualization of the $k$-nearest-neighbor (kNN) spatial graph generated from tbular dataset around the Emory area. Blue dots represent POIs and red lines indicate edges connecting each POI to its ten nearest neighbors based on geodesic distance (k=10). \pymapcopyright}
	\label{fig:graph_example_emory}
\end{figure}
Figure \ref{fig:graph_example_emory} illustrates the spatial point-graph generated from the integrated tabular dataset. Each blue node represents a POI, and each red line denotes an edge connecting a POI to its ten geographically nearest neighbors, as determined using the k-nearest-neighbor (kNN) algorithm. As shown, some clusters are disconnected from the main graph, reflecting a natural outcome of the kNN-based connectivity, where isolated regions may form independent subgraphs. In contrast, denser urban areas such as downtown and midtown exhibit a higher edge density, while predominantly residential or low-business regions contain fewer or no nodes in the graph.

\begin{table}[t]
	\centering
	\caption{Example subset of $k$-nearest-neighbor connections and distances between Foursquare–OSM POIs.}
	\label{tab:graph_sample}
	\begin{tabular}{|l|l|c|}
		\hline
		\textbf{fsq\_place\_id\_source} & \textbf{fsq\_place\_id\_destination} & \textbf{distance\_m} \\ \hline
		4ec132eb7ee54e4cd348d18b        & 4c48c4076594be9a5fba2e24             & 313.03               \\ \hline
		4ec132eb7ee54e4cd348d18b        & 5097f4fe4b9033ce2157b74d             & 347.70               \\ \hline
		4ec132eb7ee54e4cd348d18b        & 4c1e90a4eac020a18aec49c2             & 377.12               \\ \hline
		4ec132eb7ee54e4cd348d18b        & 0842aab84b994f9990b21a4d             & 385.50               \\ \hline
		4ec132eb7ee54e4cd348d18b & 4be835d4d837c9b60d05a506 & 378.52 \\ \hline
		4ec132eb7ee54e4cd348d18b & b6184fc4efce46d3f37e756b & 393.30 \\ \hline
		% 4ec132eb7ee54e4cd348d18b & 5167034be4b0008b2422fe05 & 382.10 \\ \hline
		% 4ec132eb7ee54e4cd348d18b & 5bcbca728a4cf5002c1cda90 & 405.55 \\ \hline
		% 4ec132eb7ee54e4cd348d18b & 56ba7cc9498e54e145848785 & 424.98 \\ \hline
		% 4ec132eb7ee54e4cd348d18b & 4dcac4067d8bc0c0b8854a8d & 436.20 \\ \hline
		...                             & ...                                  & ...                  \\ \hline
	\end{tabular}
\end{table}

Table \ref{tab:graph_sample} presents a sample subset of pairwise distances between POIs, where each node is connected to its \( N \) nearest neighbors. As shown, the graph data include only three fields: the source place identifier, the destination place identifier, and the distance between them (in meters). The distances were computed using a PostGIS function that calculates great-circle distances in meters rather than degrees, providing a consistent and meaningful global scale. Using meters ensures that edge weights are directly interpretable and avoids distortions that can occur when using degrees, which vary with latitude (e.g., between the equator and the poles). We note that a single \texttt{fsq\_place\_id\_source} may correspond to multiple rows in the tabular dataset. This occurs when a single Foursquare place is matched to more than one OSM location with sufficiently high name-similarity scores. As a result, a single node in the graph can represent multiple candidate matches in the tabular data. To maintain transparency and flexibility, we include all such records in the dataset and retain their corresponding place identifiers. This design allows users to apply their own filtering criteria or tie-breaking rules—such as selecting the record with the highest similarity score or the smallest spatial distance—based on the requirements of their specific analyses. An alternative solution would be to assign a new unique identifier to each record and use it as the primary key. However, this approach would create multiple distinct identifiers for what is essentially the same location, potentially leading to ambiguity and misinterpretation in the graph representation. To avoid this issue, we preserved the original \texttt{fsq\_place\_id} as the node identifier, ensuring that all records referring to the same place remain linked, while still allowing users to manage duplicate matches according to their analysis needs.

\section*{Technical Validation}
To ensure data quality and correctness of the integration process, multiple verification checks were applied at each stage of the pipeline described in the Methods section.  
In this section, we present an in-depth validation and comparison against external sources to verify the accuracy, consistency, and reliability of the integrated dataset. 

\begin{figure}[ht]
	\centering
	\begin{subfigure}{0.495\linewidth}
		\centering
		\includegraphics[width=\linewidth]{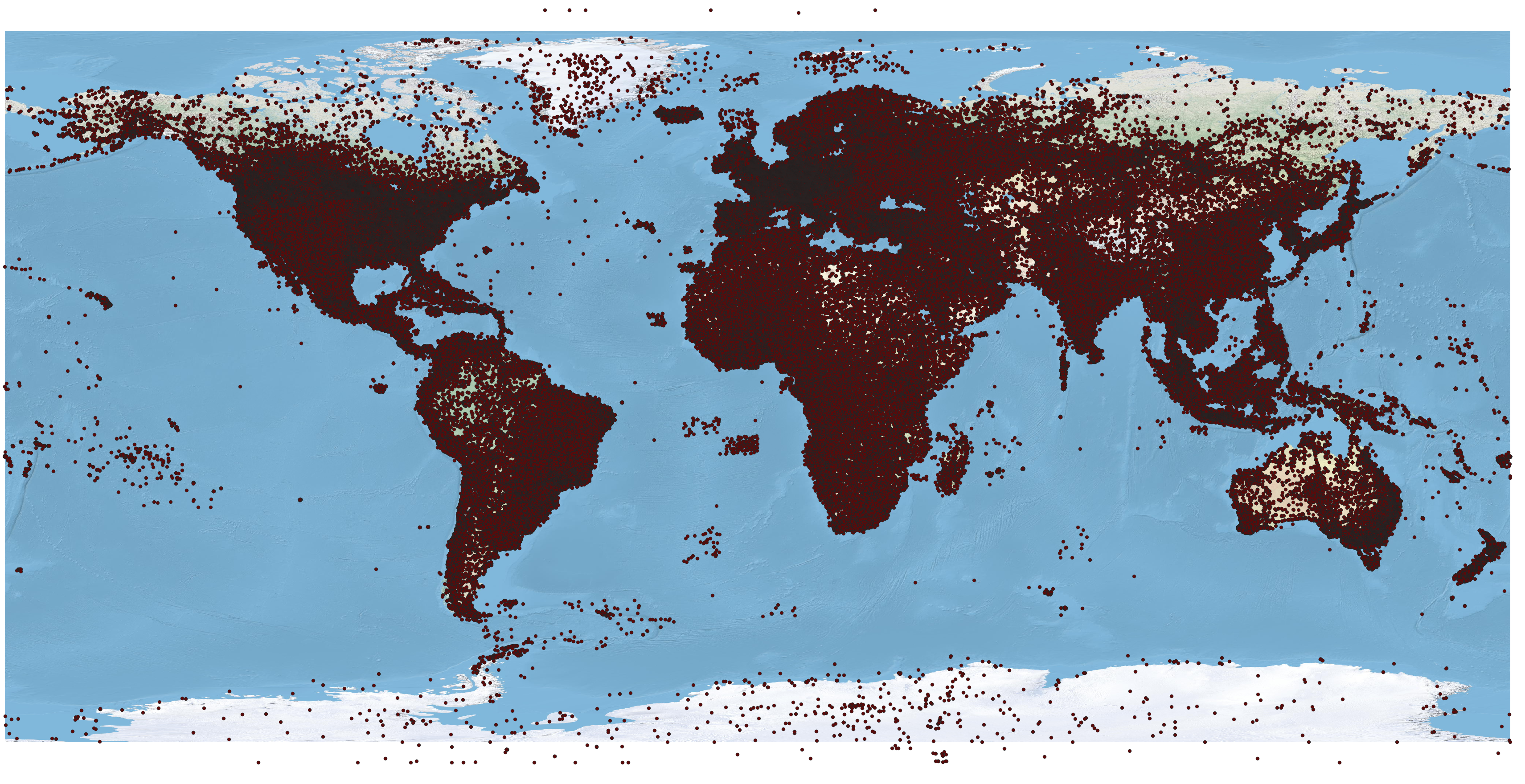}
		\caption{Foursquare POIs show broad spatial coverage, including entries in sparsely populated or uninhabited regions.  }
		\label{fig:fsq_overview}
	\end{subfigure}
	\hfill
	\begin{subfigure}{0.495\linewidth}
		\centering
		\includegraphics[width=\linewidth]{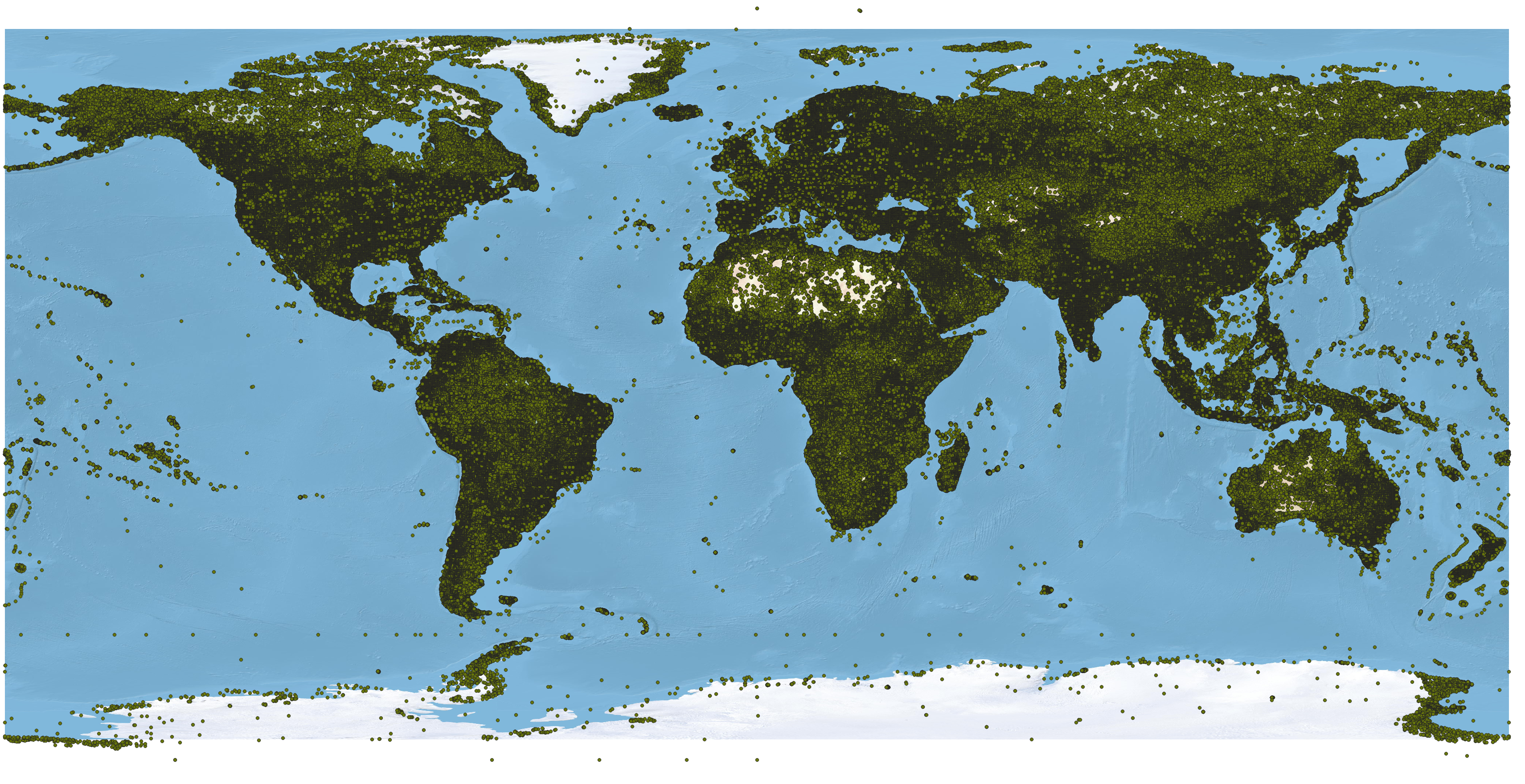}
		\caption{OSM POIs exhibit denser clustering, primarily in areas where contributors have recorded detailed map data.  }
		\label{fig:osm_overview}
	\end{subfigure}
	\hfill
	\begin{subfigure}{0.49\linewidth}
		\centering
		\includegraphics[width=1\linewidth]{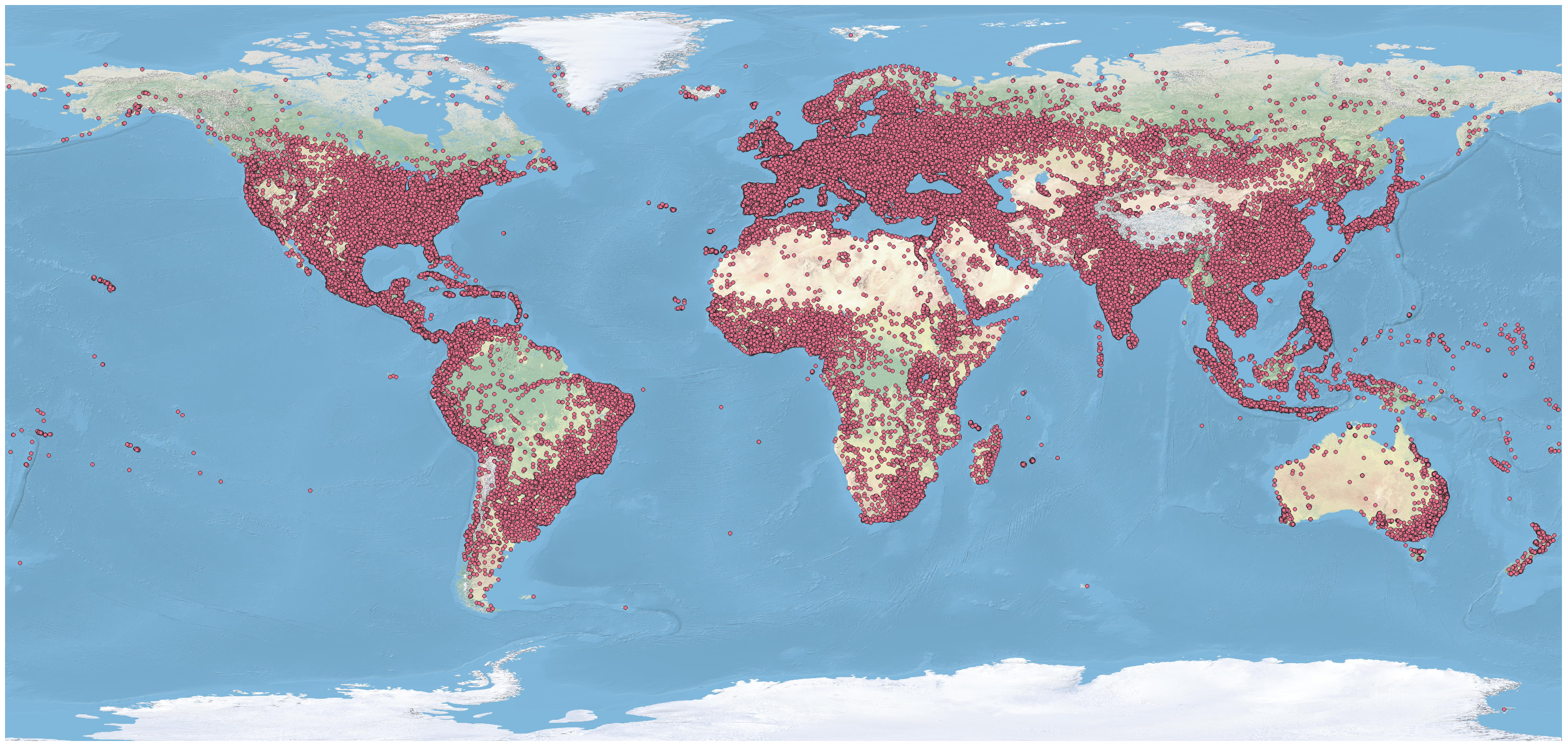}
		\caption{City population distribution (population $>$ 1,000) illustrates expected regions of human settlement and activity}
		\label{fig:population_overview}
	\end{subfigure}
	\hfill
	\begin{subfigure}{0.49\linewidth}
		\centering
		\includegraphics[width=1\linewidth]{fsq_osm_coordinates_lev2m.png}
		\caption{\datasetname{}, filtered by a Levenshtein name-similarity threshold greater than 0.5}
		\label{fig:fsq_osm_overview}
	\end{subfigure}
\caption{Visualization of Foursquare POIs, OSM points, city population distribution, and the integrated \datasetname{} dataset. The \datasetname{} dataset aligns closely with populated regions, demonstrating enhanced spatial accuracy and semantic consistency. \qgismapcopyright}

	\label{fig:overview_comparison}
\end{figure}

\subsection*{Comparison to City Population}

\reffig{fig:overview_comparison} provides a visual comparison of \datasetname{} with other data sources. As shown in \reffig{fig:fsq_overview}, the Foursquare data visualization indicates that POIs are broadly distributed across all regions, including areas that are likely uninhabited. In contrast, \reffig{fig:osm_overview} illustrates the dense clustering of OSM POIs, which generally correspond to areas that users have physically visited and mapped. The peer-reviewed and community-validated nature of OSM further enhances the accuracy and reliability of these locations.
To further validate our results, we visualize city populations in \reffig{fig:population_overview}, highlighting cities with populations exceeding 1,000—regions where POI presence is expected to be higher. Finally, \reffig{fig:fsq_osm_overview} presents the output of our integration approach, where results were filtered using a Levenshtein name-similarity threshold greater than 0.5. The strong spatial correlation between \datasetname{} and the population map demonstrates that \datasetname{} provides a realistic representation of human activity patterns across populated regions.
In other words, our approach effectively removes noise by retaining only POIs that exist in both Foursquare and OSM datasets and meet the similarity criteria, resulting in a more accurate and valid representation of real-world locations.

\begin{figure}[htbp]
	\centering
	\begin{subfigure}{0.49\linewidth}
		\centering
		\includegraphics[width=\linewidth]{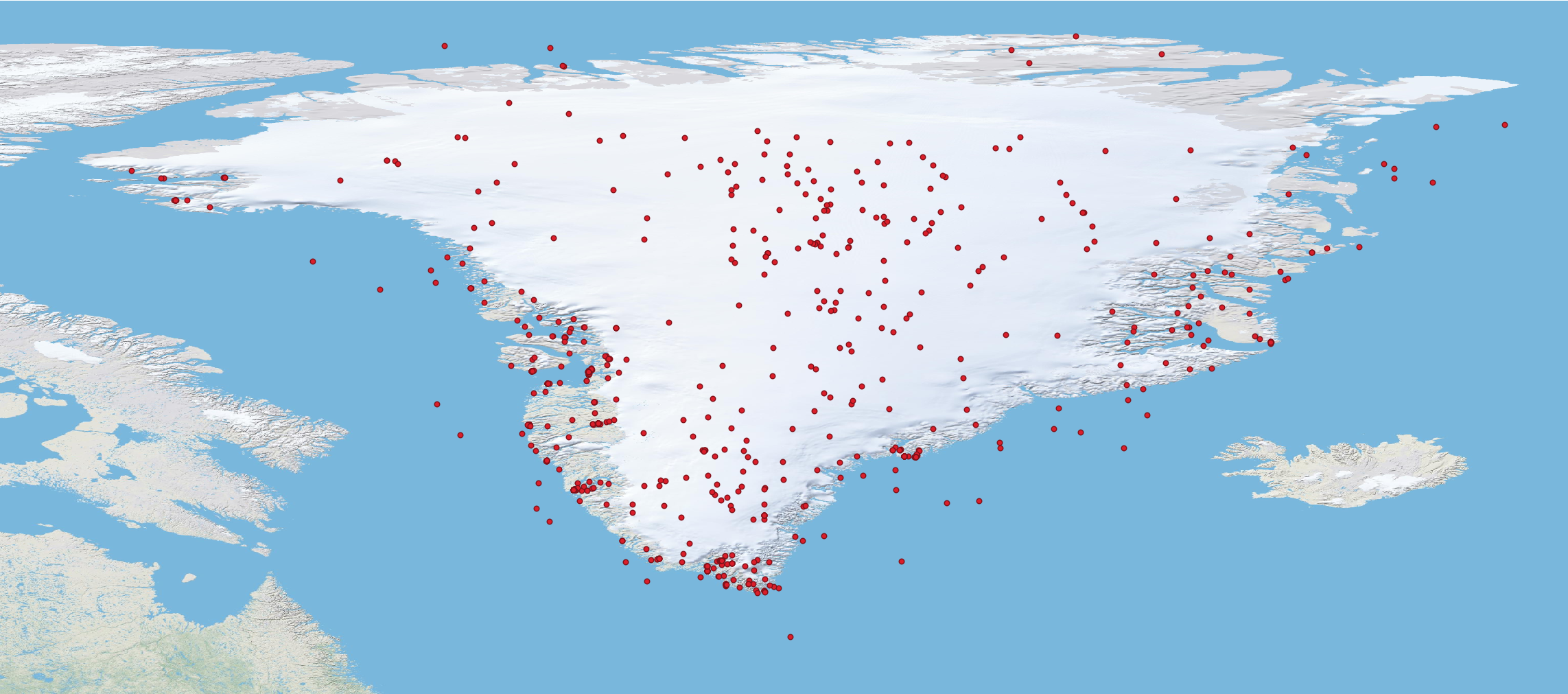}
		\caption{Foursquare POIs appear sparsely scattered, including regions with little or no permanent population.}
		\label{fig:fsq_greenland}
	\end{subfigure}
	\hfill
	\begin{subfigure}{0.49\linewidth}
		\centering
		\includegraphics[width=\linewidth]{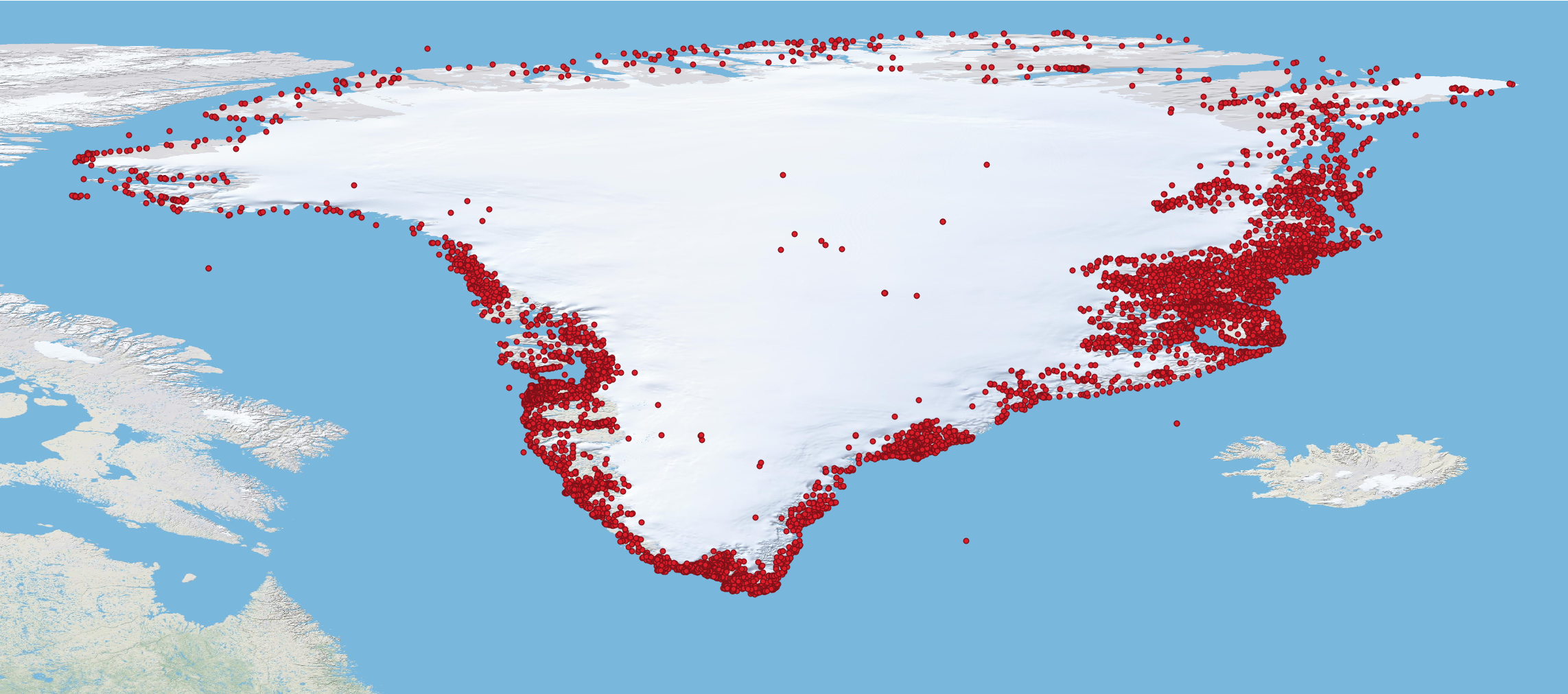}
		\caption{ OSM POIs show dense clustering along coastal regions but include entries in uninhabited areas}
		\label{fig:osm_greenland}
	\end{subfigure}
	\begin{subfigure}{0.49\linewidth}
		\centering
		\includegraphics[width=\linewidth]{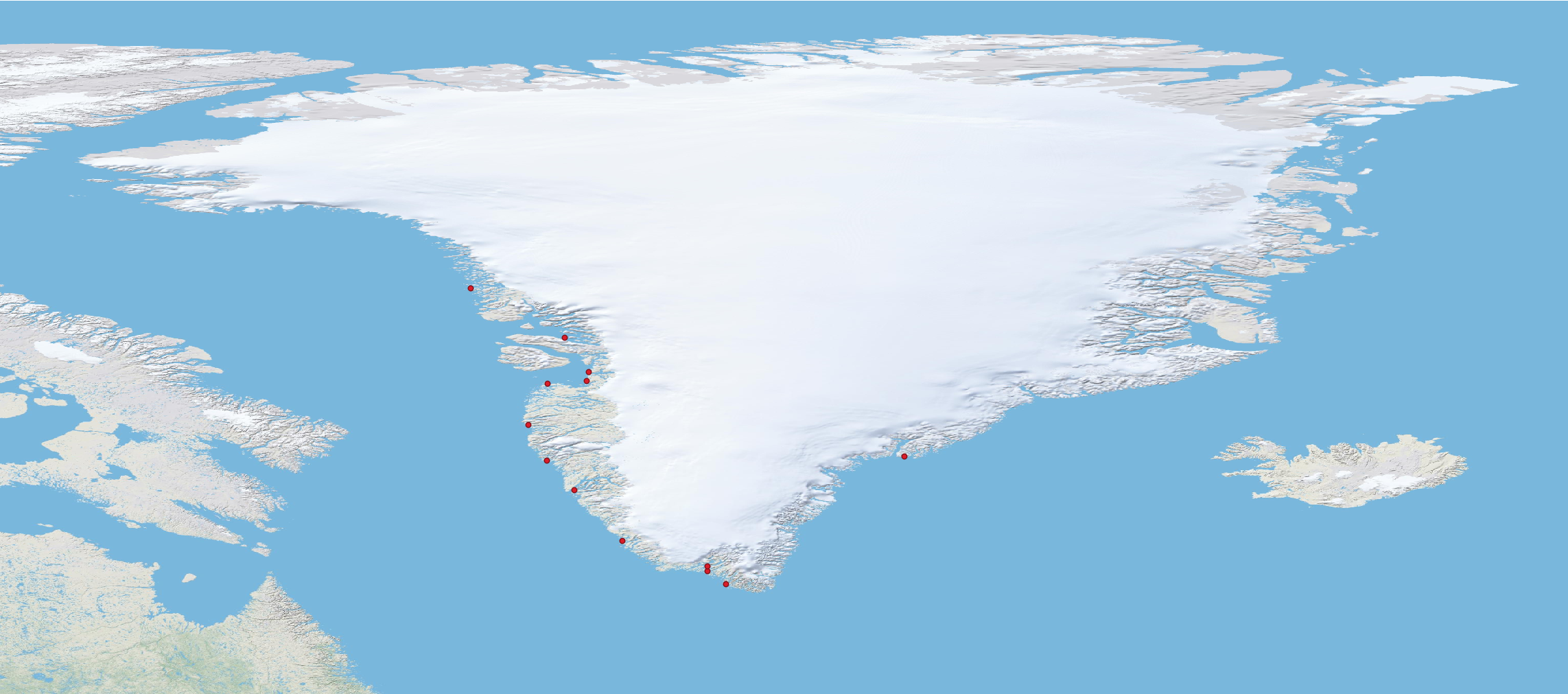}
		\caption{City population distribution (population $>$ 1,000) highlights the few populated settlements in Greenland.}
		\label{fig:population_greenland}
	\end{subfigure}
	\hfill
	\begin{subfigure}{0.49\linewidth}
		\centering
		\includegraphics[width=\linewidth]{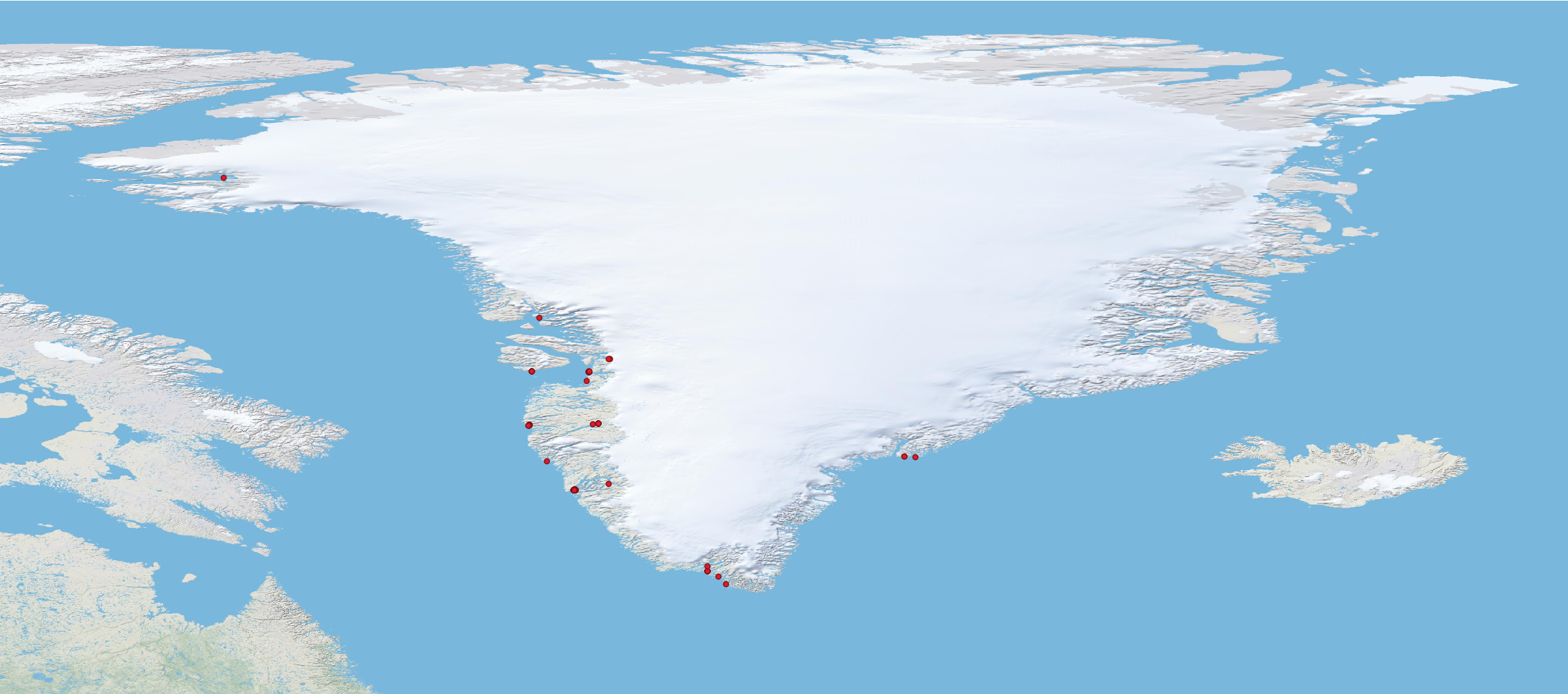}
		\caption{   \datasetname{} dataset  filtered by Levenshtein name-similarity $>$ 0.5, aligns closely with populated regions}
		\label{fig:fsq_osm_greenland}
	\end{subfigure}
	\caption{Comparison of POI distributions in Greenland across different data sources. }
	\label{fig:greenland_comparison}
\end{figure}

\subsection*{Greenland Case Study}

Manually examining every point in the dataset is not feasible; therefore, we conducted a focused case study to assess data quality. We selected Greenland as the study region because it is sparsely populated yet contains several notable points of human activity and businesses. Moreover, its POI distributions in Foursquare, OSM, population data, and \datasetname{} showed significant variation, making it a strong candidate for validation.  

\reffig{fig:greenland_comparison} presents this case study, highlighting how our integration approach refines noisy POI datasets. In \reffig{fig:fsq_greenland}, Foursquare POIs appear sparsely scattered across the island, including in uninhabited regions. In contrast, \reffig{fig:osm_greenland} shows dense clustering along the coasts but also includes numerous points in areas without permanent populations, likely resulting from automated or crowd-sourced mapping artifacts. \reffig{fig:population_greenland} visualizes cities with populations exceeding 1,000.  
The \datasetname{} subset for Greenland, shown in \reffig{fig:fsq_osm_greenland}, includes only POIs with a Levenshtein name-similarity score greater than 0.5. This filtering criterion effectively minimizes false positives and results in a spatial distribution that closely aligns with populated coastal regions. The strong correspondence between \datasetname{} and known inhabited areas demonstrates that the proposed filtering and integration methodology yields a realistic, accurate, and reliable representation of true human settlements.

The visualization provided in \reffig{fig:greenland_comparison} demonstrates the overall validity of \datasetname{}. To further evaluate the dataset, we performed a manual validation of 150 POIs: 50 from Foursquare, 50 from OSM, and 50 from \datasetname{}. Each location was cross-verified using Google Maps, and when the site could not be identified, we consulted additional sources such as Wikipedia (\url{https://wikipedia.com}), Google Search (\url{https://google.com}), and Mapcarta (\url{https://mapcarta.com/}) to obtain contextual or supporting evidence.

The results of this manual validation are summarized in \reftab{tab:greenland_sample_fsq}, \reftab{tab:greenland_sample_osm}, and \reftab{tab:greenland_sample_fsq_osm}. Overall, the majority of Foursquare POIs lacked verifiable real-world locations; despite extensive searches, many entries appeared to be synthetic or mislabeled. In contrast, OSM contained more physically realistic coordinates, yet many points corresponded to generic or unverified map features such as islands, bays, or geographic formations rather than functional POIs. 
\datasetname{}, however, produced a curated subset of POIs that corresponded to real, identifiable locations with accurate names and coordinates. Most entries in \datasetname{} were confirmed through Google Maps or other sources, indicating high spatial and semantic validity. This demonstrates that our integration and filtering pipeline effectively eliminates non-existent or ambiguous POIs while retaining accurate and meaningful locations suitable for spatial and behavioral analysis.

\paragraph{Foursquare sample (\reftab{tab:greenland_sample_fsq}).}
The Foursquare subset contains a mix of verifiable venues (e.g., \emph{Restaurant Nasaasaaq}, \emph{Sisimiut Tandklinik}, \emph{Brugseni}) and many entries that are unverifiable, out of region, or semantically non-POI tokens. Notable issues include abstract or cultural terms (\emph{Christmas 2012}, \emph{Purgatory}), planetary or foreign toponyms (\emph{Mars}, \emph{Amsterdam}, \emph{Rypefjord} in Norway), vague abbreviations (\emph{KNR}, \emph{HTX}), and generic areas rather than discrete POIs (\emph{South Greenland}, \emph{Qaasuitsup}). Several records could not be verified through any external or corroborating source, even after targeted manual searches. Overall, the sample reveals substantial noise in the Foursquare data, characterized by name inflation, redundant entries, and semantic drift away from true ground-truth venues.

Specifically, among the 50 Foursquare records examined, 11 correspond to real locations, 1 represents a permanently closed site, and 38 do not correspond to any verifiable real-world location. Of the verified real locations, 10 qualify as actual points of interest (POIs) such as businesses or facilities. Notably, the entry “Northern International Crypto Kingdom’S” appears twice in the dataset—this duplication reflects distinct Foursquare place IDs (5f5b74765148947f3da6f71b and 5f5b8389fe774e63aabdfbce) assigned to identically named records.

\paragraph{OSM sample (\reftab{tab:greenland_sample_osm}).}
The OSM subset is geographically coherent but skewed toward natural and administrative features rather than consumer POIs. Many entries are islands, bays, capes, lakes, straits, glaciers, or municipalities (e.g., \emph{Innaarsulik}, \emph{Nordfjord}, \emph{Kakivfait}, \emph{Avannaata Kommunia}), which are appropriate cartographic features but not business destinations. There is a healthy presence of genuine facilities and businesses (\emph{Spar}, \emph{Restaurant Ulo}, \emph{Hotel Arctic}, \emph{Royal Greenland}), and occasional duplicates or multi-tag representations for the same landmark (\emph{Knud Rasmussen sculpture} as both \texttt{viewpoint} and \texttt{memorial}). Ambiguous points such as \emph{N 90} drinking water and parking laybys illustrate minor tagging inconsistencies. In sum, OSM provides realistic coordinates but often for non-venue entities.

For the OSM sample, 44 entries correspond to real locations, 3 represent real-world geographic areas (such as bays, islands, or glaciers), and 3 could not be verified as real-world locations. Among the verified records, 11 correspond to identifiable points of interest (POIs) such as businesses or public facilities. Notably, Avannaata Kommunia appears twice in the dataset, with OSM IDs 405944737 and 405944738, both categorized as amenity/townhall. Similarly, Ikerasaarsuk occurs twice, with OSM IDs 413832375 and 413833463, both labeled as natural/strait. These repetitions represent legitimate multiple mappings within OSM rather than data duplication errors.

\paragraph{\datasetname{} sample (\reftab{tab:greenland_sample_fsq_osm}).}
The matched table yields high-confidence, real venues with small, interpretable positional offsets rather than semantic noise. Most pairs are legitimate places verified in maps or imagery (\emph{Nuuk Kunstmuseum}, \emph{Café Inuk}, \emph{Ilisimatusarfik}, \emph{STARK Sisimiut}, \emph{Hotel Kulusuk}). A common pattern is sub-building to street-side displacement where the FSQ point lies just outside the structure and OSM lies on the building footprint (\emph{Fitness GL}, \emph{Hotel Sisimiut}, \emph{Blue Café}). The table also captures operational status and context (\emph{SparQuick} permanently closed; \emph{Nuugaatsiaq} now abandoned), and highlights occasional name variants or multiple OSM IDs for the same venue (\emph{Hotel Disko Island}). One clear geocoding error (\emph{Ilulissat Vandrehjem} matched to a house due to a misplaced FSQ point) underscores the value of dual-source matching. Overall, the matched set preserves genuine, mappable venues with building-level offsets typical of heterogeneous geocoders, supporting its suitability for spatial analysis.

For the \datasetname{} sample, 48 entries correspond to real locations, 2 represent permanently closed sites, and 2 correspond to broader areas or villages. Among these, 45 out of 48 verified entries are valid points of interest (POIs). The entry Suloraq illustrates a unique case—its Foursquare record corresponds to an apartment complex, while the OSM data include both Suloraq 11 (a building) and Suloraq (a nearby statue). Because the primary match represents residential housing rather than a commercial or functional POI, Suloraq was excluded from the POI count. Similarly, Ilulissat Vandrehjem corresponds to a private house rather than a POI.
Additionally, duplicate mappings were identified where Brugseni was matched to two OSM records (180978036 and 1913382879) and Hotel Disko was mapped to two OSM records (715178079 and 4399888527), both representing legitimate but redundant OSM entries of the same establishment.

\section*{Usage Note}

The \datasetname{} dataset offers a unified, high-quality view of global points of interest (POIs) by integrating Foursquare and OpenStreetMap (OSM) data through spatial and semantic alignment. This integration enables a wide range of downstream research and analytical applications across geospatial, data science, and urban informatics domains. Below, we describe key potential use cases.

\paragraph{1. Geographic Knowledge Graph Construction.}
The dataset provides a foundation for building geographic knowledge graphs that link semantically and spatially related POIs across heterogeneous data sources. Each record includes identifiers, metadata, and similarity metrics that facilitate relationship extraction and entity resolution. This structure enables machine-readable representations of urban spaces, supporting tasks such as graph-based reasoning, spatial relationship modeling, and semantic query expansion.

\paragraph{2. Location-Based Clustering and Classification.}
The spatial coordinates and category labels of POIs can be used for unsupervised clustering and classification analyses. Researchers can identify spatial clusters of economic activity, cultural landmarks, or service facilities and classify regions based on POI density and diversity. The combination of spatial and semantic similarity fields allows advanced methods such as DBSCAN, k-means, or hierarchical clustering to reveal patterns of urban structure and human activity.

\paragraph{3. Semantic Enrichment of Commercial POI Data.}
\datasetname{} enables the semantic enrichment of commercial or proprietary POI datasets using OSM’s open and extensible tagging framework. Users can cross-link business listings with OSM attributes such as amenities, building types, or accessibility indicators, improving data completeness and interoperability. This feature is especially useful for data fusion in urban analytics, retail mapping, and smart city applications.

\paragraph{4. Urban and Regional Planning Analysis.}
The integrated dataset can be employed to assess spatial accessibility, infrastructure distribution, and land-use balance across cities or regions. Planners can identify underserved neighborhoods, evaluate commercial diversity, and monitor urban growth trends over time by leveraging the dataset’s standardized geographic and categorical attributes.

\paragraph{5. Mobility and Human Behavior Modeling.}
Because the dataset accurately reflects real-world activity centers, it is suitable for mobility modeling, travel demand forecasting, and agent-based simulations. Each verified POI represents a potential anchor for human movement, allowing integration into population mobility models or epidemic simulations where realistic behavioral and spatial dynamics are required.

\paragraph{6. Data Quality Assessment and Benchmarking.}
\datasetname{} can serve as a benchmark for evaluating geocoding accuracy, record linkage algorithms, or POI data quality assessment methods. Its verified matches between Foursquare and OSM records, coupled with name similarity metrics and distance fields, allow researchers to systematically test and calibrate entity-matching pipelines or evaluate spatial accuracy in noisy or incomplete datasets.

\paragraph{7. Network and Graph-Based Analyses.}
The provided graph representation of POIs, where nodes represent locations and edges denote k-nearest spatial relationships, enables graph-theoretic analyses such as centrality, clustering coefficients, and community detection. These analyses can reveal patterns of urban connectivity, identify local hubs of activity, and support transportation or accessibility modeling.

Together, these use cases highlight \datasetname{} as a robust and versatile resource for spatial data integration, geoinformatics, and computational urban studies. Its reproducible design and global coverage make it suitable for both methodological research and applied spatial analysis at multiple geographic scales.

\bibliography{main}
\section*{Code Availability}

The complete codebase used for data collection, preprocessing, spatial integration, and export is publicly available at:
\githuburl

The repository includes:

Scripts for downloading and parsing Foursquare and OSM data.

SQL and Python code for spatial processing and matching.

Tools for generating graph representations and exporting structured CSVs.

Configuration files and reproducibility documentation.

\section*{Acknowledgments}
This research was supported by the National Science Foundation (NSF) under Award No. 2109647, \textit{Data-Driven Modeling to Improve Understanding of Human Behavior, Mobility, and Disease Spread}.

\section*{Author contributions statement}

HA collected and processed the data and prepared the draft manuscript. MH generated the graph data and also contributed to the draft manuscript. AZ designed and conducted the analysis and participated in writing the draft manuscript.

\section*{Competing interests}
The authors declare no competing interests.

\begin{table}
\centering
	\caption{Manual validation of a random sample of 50 \textbf{Foursquare} records from Greenland. Validation was performed using Google Maps and other online sources to determine whether each listed place exists and, if so, whether it represents an actual business.}
	\label{tab:greenland_sample_fsq}
	\footnotesize
	\begin{tabular}{|m{0.26\linewidth} | m{0.4\linewidth} | C{0.07\linewidth} |m{0.17\linewidth} |}
		\hline
		\textbf{fsq\_name }                                & \textbf{fsq\_category\_labels}                                                                 & \textbf{Business} & \textbf{Observation} \\ \hline
		Akiki                                     & ['Retail > Food and Beverage Retail > Grocery Store']                                 & YES          & Real Location                        \\ \hline
		Brugseni                                  & ['Retail > Food and Beverage Retail > Supermarket']                                   & YES          & Real Location                        \\ \hline
		STARK Sisimiut                            & ['Retail > Hardware Store']                                                           & YES          & Real Location                        \\ \hline
		Restaurant Nasaasaaq                      & ['Dining and Drinking > Restaurant']                                                  & YES          & Real Location                        \\ \hline
		Nuuk Kunstmuseum                          & ['Arts and Entertainment > Museum > Art Museum']                                      & YES          & Real Location                        \\ \hline
		Sisimiut Tandklinik                       & ['Health and Medicine > Dentist']                                                     & YES          & Real Location                        \\ \hline
		Sisimiut Harbour                          & ['Landmarks and Outdoors > Harbor or Marina']                                         & YES          & Real Location                        \\ \hline
		The Culture Centre Taseralik              & ['Arts and Entertainment > Performing Arts Venue > Concert Hall']                     & YES          & Real Location                        \\ \hline
		Nana's Thai Takeaway                      & ['Dining and Drinking > Restaurant > Asian Restaurant > Thai Restaurant']             & YES          & Real Location                        \\ \hline
		Paamaap Kuua 13-307                       & ['Community and Government > Assisted Living']                                        & YES          & Real Location                        \\ \hline
		sermeq kujataleq                          & ['Landmarks and Outdoors > Scenic Lookout']                                           & NO           & Real Location                        \\ \hline
		Spar Quick                                & ['Retail > Miscellaneous Store']                                                      & N/A          & Real Location (Closed)               \\ \hline
		Qaasuitsup                                & ['Landmarks and Outdoors > States and Municipalities > Town']                         & N/A          & NO Real Location                        \\ \hline
		South Greenland                           & ['Landmarks and Outdoors > Island']                                                   & N/A          & NO Real Location                        \\ \hline
		Nuuk                                      & ['Travel and Transportation > Transport Hub > Airport']                               & N/A          & NO Real Location                        \\ \hline
		Groelandia                                & ['Landmarks and Outdoors > Field']                                                    & N/A           & NO Real Location                        \\ \hline
		Biblioteket - The Library                 & ['Community and Government > Library']                                                & N/A           & NO Real Location                        \\ \hline
		Christmas 2012                            & ['Arts and Entertainment']                                                            & N/A           & NO Real Location                        \\ \hline
		Purgatory                                 & ['Landmarks and Outdoors > Campground']                                               & N/A           & NO Real Location                        \\ \hline
		48 GU 556                                 & ['Travel and Transportation > Road']                                                  & N/A           & NO Real Location                        \\ \hline
		Mars                                      & ['Travel and Transportation > Rest Area']                                             & N/A           & NO Real Location                        \\ \hline
		Amsterdam                                 & ['Landmarks and Outdoors > Mountain']                                                 & N/A           & NO Real Location                        \\ \hline
		KNR                                       & ['Business and Professional Services > TV Station']                                   & N/A           & NO Real Location                        \\ \hline
		Fit Life Fitness                          & ['Sports and Recreation']                                                             & N/A           & NO Real Location                        \\ \hline
		Antartika                                 & ['Landmarks and Outdoors > Mountain']                                                 & N/A           & NO Real Location                        \\ \hline
		Marlin Shadow Master Lloyd Waters         & ['Landmarks and Outdoors > Park > National Park']                                     & N/A           & NO Real Location                        \\ \hline
		Tall Junk 45 Kite                         & ['Landmarks and Outdoors > Park > National Park']                                     & N/A           & NO Real Location                        \\ \hline
		Big Aster Dong Water Over Lloyd Forest 67 & ['Landmarks and Outdoors > Park > National Park']                                     & N/A           & NO Real Location                        \\ \hline
		Davis Strait                              & ['Landmarks and Outdoors > Other Great Outdoors']                                     & N/A           & NO Real Location                        \\ \hline
		Last Intermediate National Kartel         & ['Landmarks and Outdoors > Park > National Park']                                     & N/A           & NO Real Location                        \\ \hline
		Mine Of New Youth Making Erosion          & ['Landmarks and Outdoors > Park > National Park']                                     & N/A           & NO Real Location                        \\ \hline
		Party Hard On the Boulevard               & ['Community and Government > Spiritual Center > Temple']                              & N/A           & NO Real Location                        \\ \hline
		Northern International Crypto Kingdom'S   & ['Landmarks and Outdoors > Park > National Park']                                     & N/A           & NO Real Location                        \\ \hline %5f5b74765148947f3da6f71b
		Northern International Crypto Kingdom'S   & ['Landmarks and Outdoors > Park > National Park']                                     & N/A           & NO Real Location                        \\ \hline % 5f5b8389fe774e63aabdfbce
		Eslant Sentinel Castle                    & ['Landmarks and Outdoors > Castle']                                                   & N/A           & NO Real Location                        \\ \hline
		Afsane                                    & ['Dining and Drinking > Restaurant > Diner']                                          & N/A           & NO Real Location                        \\ \hline
		Bayramyeri Sok                            & ['Retail > Big Box Store']                                                            & N/A           & NO Real Location                        \\ \hline
		Rypefjord                                 & ['Landmarks and Outdoors > Bay']                                                      & N/A           & NO Real Location                        \\ \hline
		Polo Norte                                & ['Landmarks and Outdoors > Mountain']                                                 & N/A           & NO Real Location                        \\ \hline
		Igloo Mountain                            & ['Sports and Recreation > Snow Sports > Ski Lodge']                                   & N/A           & NO Real Location                        \\ \hline
		Griffith South Center                     & ['Landmarks and Outdoors > Structure']                                                & N/A           & NO Real Location                        \\ \hline
		Tacan                                     & ['Landmarks and Outdoors > Field']                                                    & N/A           & NO Real Location                        \\ \hline
		baghimiz                                  & ['Landmarks and Outdoors > Botanical Garden']                                         & N/A           & NO Real Location                        \\ \hline
		Haluk Kece                                & ['Event > Conference']                                                                & N/A           & NO Real Location                        \\ \hline
		Afrika                                    & ['Travel and Transportation']                                                         & N/A           & NO Real Location                        \\ \hline
		Ekvator                                   & ['Arts and Entertainment > Aquarium']                                                 & N/A           & NO Real Location                        \\ \hline
		Alpler                                    & ['Landmarks and Outdoors > Mountain']                                                 & N/A           & NO Real Location                        \\ \hline
		HTX                                       & ['Community and Government > Education > College and University > College Cafeteria'] & N/A           & NO Real Location                        \\ \hline
		Bar Aaveq                                 & ['Dining and Drinking > Bar']                                                         & N/A           & NO Real Location                        \\ \hline
		The Hut                                   & ['Landmarks and Outdoors > Scenic Lookout']                                           & N/A           & NO Real Location                        \\ \hline
	\end{tabular}
\end{table}

\begin{table}
	\caption{Manual validation of a random sample of 50 \textbf{OpenStreetMap (OSM)} records from Greenland. Validation was performed using Google Maps and other online sources to determine whether each listed place exists and, if so, whether it represents an actual business.}
	\label{tab:greenland_sample_osm}
	\footnotesize
	\begin{tabular}{| p{0.3\linewidth} | p{0.15\linewidth} | p{0.15\linewidth} | C{0.07\linewidth} | m{0.17\linewidth} |}
		\hline
		\textbf{osm\_name}                & \textbf{osm\_class} & \textbf{osm\_type}       & \textbf{Business} & \textbf{Observation} \\ \hline
		Akiki                    & shop       & supermarket     & YES          & Real Location                        \\ \hline
		Hotel Arctic             & building   & hotel           & YES          & Real Location                        \\ \hline
		Spar                     & shop       & supermarket     & YES          & Real Location                        \\ \hline
		Kangerluarsuk            & natural    & bay             & NO           & Real Location                        \\ \hline
		Innaarsulik              & place      & islet           & NO           & Real Location                        \\ \hline
		Angisunnguaq             & place      & island          & NO           & Real Location                        \\ \hline
		Kap Udkiggen             & natural    & cape            & NO           & Real Location                        \\ \hline
		Nordfjord                & natural    & bay             & NO           & Real Location                        \\ \hline
		Krumodden                & natural    & cape            & NO           & Real Location                        \\ \hline
		Pooqattaq                & place      & island          & NO           & Real Location                        \\ \hline
		Arctic Umiaq Line        & amenity    & ferry\_terminal & YES          & Real Location                        \\ \hline
		Aqqutarfik               & tourism    & museum          & YES          & Real Location                        \\ \hline
		Royal Greenland          & building   & industrial      & YES          & Real Location                        \\ \hline
		Royal Arctic Line        & building   & warehouse       & YES          & Real Location                        \\ \hline
		Polar Oil                & building   & yes             & YES          & Real Location                        \\ \hline
		Sunset Boulevard         & amenity    & fast\_food      & YES          & Real Location                        \\ \hline
		Knud Rasmussen sculpture & tourism    & viewpoint       & NO           & Real Location                        \\ \hline
		Knud Rasmussen sculpture & historic   & memorial        & NO           & Real Location                        \\ \hline
		Restaurant Ulo           & amenity    & restaurant      & YES          & Real Location                        \\ \hline
		Orpissuup Tasia          & water      & lake            & NO           & Real Location                        \\ \hline
		Butik Sara               & shop       & gift            & YES          & Real Location                        \\ \hline
		Eqalunnguit Nunaat       & place      & island          & NO           & Real Location                        \\ \hline
		Tasersuaq                & water      & lake            & NO           & Real Location                        \\ \hline
		Kangiilat                & place      & locality        & NO           & Real Location                        \\ \hline
		Apuliliip Apusiia        & natural    & glacier         & NO           & Real Location                        \\ \hline
		Qeqertarsuatsiaq         & place      & island          & NO           & Real Location                        \\ \hline
		Sermeq (2012)            & natural    & glacier         & NO           & Real Location                        \\ \hline
		Iggiaanut - Bro 1        & amenity    & parking         & NO           & Real Location                        \\ \hline
		Inalugartuut Iluat       & natural    & bay             & NO           & Real Location                        \\ \hline
		Kangiussap Paava         & natural    & bay             & NO           & Real Location                        \\ \hline
		Aarfit Timaat            & place      & island          & NO           & Real Location                        \\ \hline
		Kangeq                   & place      & island          & NO           & Real Location                        \\ \hline
		Amitsuarssuk             & natural    & bay             & NO           & Real Location                        \\ \hline
		Aasiaat                  & place      & island          & NO           & Real Location                        \\ \hline
		Ungoorsivik              & natural    & bay             & NO           & Real Location                        \\ \hline
		Eqalunnguit              & natural    & bay             & NO           & Real Location                        \\ \hline
		Majooq                   & place      & island          & NO           & Real Location                        \\ \hline
		Niaqornaq                & place      & islet           & NO           & Real Location                        \\ \hline
		Kakivfait                & natural    & strait          & NO           & Real Location                        \\ \hline
		Anarusuk                 & place      & island          & NO           & Real Location                        \\ \hline
		Eqalugaarsuit            & natural    & bay             & NO           & Real Location                        \\ \hline
		Ikerasaarsuk             & natural    & strait          & N/A           & Real Location                 \\ \hline % (osm\_id=413832375) 
		Ikerasaarsuk             & natural    & strait          & N/A           & Real Location                  \\ \hline % (osm\_id=a413833463) 
		Nuuk Imeq A/S            & building   & industrial      & N/A           & Real Location (Closed)                       \\ \hline
		Avannaata Kommunia       & amenity    & townhall        & N/A           & AREA                  \\ \hline %  (osm\_id=405944737)
		Avannaata Kommunia       & amenity    & townhall        & N/A           & AREA                   \\ \hline % (osm\_id=405944738) 
		Pinngorsuaq              & place      & neighbourhood   & N/A           & AREA                  \\ \hline
		Fiskefabrik              & shop       & seafood         & N/A           & NO Real Location                        \\ \hline
		N 90 E                   & amenity    & drinking\_water & N/A           & NO Real Location                        \\ \hline
		N 90                     & amenity    & drinking\_water & N/A           & NO Real Location                        \\ \hline
	\end{tabular}
\end{table}

\begin{table}
	\caption{Manual validation of a random sample of 50 \textbf{\datasetname{}} records from Greenland. Validation was performed using Google Maps and other online sources to determine whether each listed place exists and, if so, whether it represents an actual business.}
	\label{tab:greenland_sample_fsq_osm}
	\footnotesize
	\begin{tabular}{| m{0.25\linewidth} | m{0.25\linewidth} | m{0.1\linewidth} | C{0.07\linewidth} | m{0.17\linewidth} |}
		\hline
		\textbf{fsq\_name}                                          & \textbf{osm\_name}                                   & \textbf{osm\_type}       & \textbf{Business} & \textbf{Observation} \\ \hline
		Nuuk Kunstmuseum                                   & Nuuk Kunstmuseum                            & museum          & YES          & Real Location                        \\ \hline
		Fitnessgl                                          & Fitness GL                                  & fitness\_centre & YES          & Real Location                        \\ \hline
		Salon Nauja                                        & Salon Nauja                                 & hairdresser     & YES          & Real Location                        \\ \hline
		Arktisk Kommando                                   & Arktisk Kommando                            & government      & YES          & Real Location                        \\ \hline
		Comby A/S                                          & Comby                                       & it              & YES          & Real Location                        \\ \hline
		Nukissiorfiit                                      & Nukissiorfiit                               & government      & YES          & Real Location                        \\ \hline
		Café Inuk                                          & Café Inuk                                   & cafe            & YES          & Real Location                        \\ \hline
		Ilisimatusarfik                                    & Ilisimatusarfik                             & university      & YES          & Real Location                        \\ \hline
		Grønlands Naturinstitut                            & Grønlands Naturinstitut                     & parking         & YES          & Real Location                        \\ \hline
		Nuuk Golfklub                                      & Nuuk Golf Club                              & yes             & YES          & Real Location                        \\ \hline
		Suloraq                                            & Suloraq 11                                  & apartments      & NO           & Real Location                        \\ \hline
		Suloraq                                            & Suloraq                                     & artwork         & NO           & Real Location                        \\ \hline
		Pisiffik Qinngorput                                & Pisiffik Qinngorput                         & supermarket     & YES          & Real Location                        \\ \hline
		Maniitsoq Harbour                                  & Maniitsoq                                   & ferry\_terminal & YES          & Real Location                        \\ \hline
		Hotel Søma Sisimiut                                & Hotel SØMA                                  & hotel           & YES          & Real Location                        \\ \hline
		Hotel Sisimiut                                     & Hotel Sisimiut                              & hotel           & YES          & Real Location                        \\ \hline
		STARK Sisimiut                                     & STARK Sisimiut                              & hardware        & YES          & Real Location                        \\ \hline
		Brugseni                                           & Brugseni                                    & supermarket     & YES          & Real Location                        \\ \hline % (osm\_id=180978036) 
		Brugseni                                           & Brugseni                                    & supermarket     & YES          & Real Location                        \\ \hline % (osm\_id=1913382879) 
		Nana's Thai Takeaway                               & N/Anas Thai Take away                       & fast\_food      & YES          & Real Location                        \\ \hline
		Pub Raaja                                          & Pub Raaja                                   & pub             & YES          & Real Location                        \\ \hline
		Panorama                                           & Panorama Hostel                             & hotel           & YES          & Real Location                        \\ \hline
		Hotel Disko                                        & Hotel Disko Island                          & hotel           & YES          & Real Location                        \\ \hline % (osm\_id=715178079) 
		Hotel Disko                                        & Hotel Disko Island                          & hotel           & YES          & Real Location                        \\ \hline % (osm\_id=4399888527) 
		Cafe Blue                                          & Blue Café                                   & cafe            & YES          & Real Location                        \\ \hline
		Skansen - Dit hjem                                 & Skansen Dit Hjem                            & chalet          & YES          & Real Location                        \\ \hline
		Hotel Kulusuk                                      & Hotel Kulusuk                               & hotel           & YES          & Real Location                        \\ \hline
		ammassalik museum                                  & Ammassalik Museum                           & museum          & YES          & Real Location                        \\ \hline
		Hotel Angmassalik                                  & Angmagssalik Hotel                          & hotel           & YES          & Real Location                        \\ \hline
		Efterskole Villads Villadsen                       & Efterskole Villads Villadsen                & school          & YES          & Real Location                        \\ \hline
		Cafe Nuka                                          & Cafe Nuka                                   & cafe            & YES          & Real Location                        \\ \hline
		Hangout Bistro                                     & Hangout Bistro                              & fast\_food      & YES          & Real Location                        \\ \hline
		Best Western Plus Hotel Ilulissat                  & Best Western Plus Hotel Ilulissat           & hotel           & YES          & Real Location                        \\ \hline
		Hotel Icefiord                                     & Hotel Icefiord                              & hotel           & YES          & Real Location                        \\ \hline
		Akiki                                              & Akiki                                       & supermarket     & YES          & Real Location                        \\ \hline
		The Glacier Shop                                   & Glacier Shop                                & gift            & YES          & Real Location                        \\ \hline
		Cafennguaq                                         & Cafénnguaq                                  & cafe            & YES          & Real Location                        \\ \hline
		Hotel Arctic Igloos                                & Hotel Arctic                                & hotel           & YES          & Real Location                        \\ \hline
		Ilulissat Vandrehjem                               & Vandrehjem                                  & house           & NO           & Real Location                        \\ \hline
		Ilulissat Airport                                  & Ilulissat                                   & terminal        & YES          & Real Location                        \\ \hline
		Eqip Sermia / Eqip Glacier                         & Eqip Sermia                                 & viewpoint       & YES          & Real Location                        \\ \hline
		Eqi Glacier Lodge                                  & Glacier Lodge Eqi                           & camp\_site      & YES          & Real Location                        \\ \hline
		Kangerlussuaq International Science Support Center & KISS (Kangerlussuaq Intn'l Science Support) & hotel           & YES          & Real Location                        \\ \hline
		Kang Mini Marked                                   & Kang mini marked                            & supermarket     & YES          & Real Location                        \\ \hline
		Kangerlussuaq Youth Hostel / Vandrehjem            & Kangerlussuaq Vandrehjem                    & hostel          & YES          & Real Location                        \\ \hline
		Polar Lodge                                        & Polar Lodge                                 & hotel           & YES          & Real Location                        \\ \hline
		Spar Quick                                         & SparQuick                                   & convenience     & N/A           & Real Location (Closed)                         \\ \hline
		Cafe Naapiffik                                     & Cafe Naapiffik                              & cafe            & N/A           & Real Location  (Closed)                       \\ \hline
		Nuugaatsiaq (city)                                 & Nuugaatsiaq                                 & village         & N/A           & AREA (Abandoned)                        \\ \hline
		Qeqertat                                           & Qeqertat                                    & village         & N/A           & AREA                       \\ \hline
		% Sisimiut Vandrehjem                                & Sisimiut Vandrehjem                             & hostel        & YES                & YES   & fsq location outside the building, osm location next to the building on a rock     \\ \hline
		% Muskox Restaurant                                  & Restaurant Muskox                               & restaurant     & YES                & YES  &   location true    \\ \hline
		% Sondrestrom Upper Atmospheric Research Facility    & Sondrestrom Upper Atmospheric Research Facility & yes           & YES                & YES   &   location true    \\ \hline
		% Café Iluliaq	                                   & Café Iluliaq	                                 & cafe	           & YES                & YES & fsq and osm location outside the building\\ \hline
		% Ishuset                                            & Ishuset                                         & convenience   & YES                & YES   &  osm\_id=8910188849    \\ \hline
		% Ishuset                                            & Ishuset                                         & convenience   & YES                & YES   &  osm\_id=181240219    \\ \hline
	\end{tabular}
\end{table}

\end{document}